\documentclass[11pt]{article}
\usepackage[noblocks]{authblk}
\usepackage{titlesec}
\usepackage{geometry}
\usepackage[square,sort,comma,numbers]{natbib}
\usepackage{graphicx}
\usepackage[english]{babel}
\usepackage{amsmath}
\usepackage{amssymb}
\usepackage{amsfonts}
\usepackage{amsthm}
\usepackage{thmtools} 
\usepackage{thm-restate}
\usepackage{accents}
\usepackage{caption}
\usepackage{subcaption}

\usepackage[dvipsnames]{xcolor}
\definecolor{cb-black}      {RGB}{  0,   0,   0}
\definecolor{cb-blue-green} {RGB}{  0,  073,  073}
\definecolor{cb-green-sea}  {RGB}{  0, 146, 146}
\definecolor{cb-rose}       {RGB}{255, 109, 182}
\definecolor{cb-salmon-pink}{RGB}{255, 182, 119}
\definecolor{cb-purple}     {RGB}{ 73,   0, 146}
\definecolor{cb-blue}       {RGB}{ 0, 109, 219}
\definecolor{cb-lilac}      {RGB}{182, 109, 255}
\definecolor{cb-blue-sky}   {RGB}{109, 182, 255}
\definecolor{cb-blue-light} {RGB}{182, 219, 255}
\definecolor{cb-burgundy}   {RGB}{146,   0,   0}
\definecolor{cb-brown}      {RGB}{146,  73,   0}
\definecolor{cb-clay}       {RGB}{219, 209,   0}
\definecolor{cb-green-lime} {RGB}{ 36, 255,  36}
\definecolor{cb-yellow}     {RGB}{255, 255, 109}

\usepackage{tikz}
\usepackage[compat=newest]{yquant}
\usepackage[colorlinks,linkcolor=cb-burgundy,urlcolor=cb-burgundy,citecolor=cb-burgundy,destlabel]{hyperref}
\usepackage[nameinlink,capitalise]{cleveref}
\usepackage{stmaryrd}
\usetikzlibrary{external}
\usetikzlibrary{quotes}
\usetikzlibrary{calc}
\usetikzlibrary{fit}

\newcommand{\ket}[1]{|#1\rangle}

\newtheorem{theorem}{Theorem}
\counterwithin{figure}{section}
\counterwithin{table}{section}
\counterwithin{theorem}{section}

\newtheorem{problem}[theorem]{Problem}

\theoremstyle{remark}

\declaretheoremstyle[
  notefont=\mdseries, notebraces={(}{)},
  bodyfont=\normalfont,
  postheadspace=0em,
  headpunct=
]{algostyle}
\theoremstyle{algostyle}

\newcommand*{\ip}[1]{ \langle #1 \rangle } 
\newcommand*{\nrm}[1]{ \lVert #1 \rVert } 
 
\newcommand*{\Tr}{\mathrm{Tr}} 

\title{Direct U(2) approximation via repeat-until-success circuits}

\author[1]{Vadym Kliuchnikov\thanks{Current address: NVIDIA, Toronto, ON M5V 1K4, Canada}}
\affil[1]{Microsoft Quantum, Toronto, ON M5J 0E7, Canada}
\author[2]{Jendrik Brachter\thanks{J.B. worked on this paper during his internship at Microsoft Research during summer of 2018.}}
\affil[2]{Microsoft Research, Redmond, WA 98052, USA}
\author[3]{Marcus P. da Silva}
\affil[3]{Microsoft Quantum, Redmond, WA 98052, USA}

\begin{document}

\maketitle

\begin{abstract}
We show how to directly and efficiently approximate arbitrary one-qubit unitaries, bypassing 
the Euler decomposition and the magnitude approximation problem, at the cost of one ancillary qubit. 
Our technique also applies to approximating unitaries with multi-qubit gate sets such as Clifford and CS, or Clifford and CCZ, as well as to approximating orthogonal matrices using multi-qubit gate sets such as Real Clifford and CCZ. The key tools are repeat-until-success circuits, lattice-based exact synthesis algorithms, integer point enumeration in convex sets, and relative norm equations.
\end{abstract}

\section{Introduction}

Compiling quantum algorithms for fault-tolerant architectures is a fundamental challenge and an active area of research in quantum computing. 
While fault-tolerant quantum computation offers a path to scalable quantum computing,
it comes with significant constraints on the gate sets that can be implemented natively. 
Many gates used in quantum algorithms cannot be exactly implemented using fault-tolerant gate sets and must 
instead be approximated by sequences of gates from a discrete universal gate set.
This limitation is not merely practical but fundamental. There exist rigorous no-go results establishing that 
certain natural unitaries cannot be exactly synthesized over common fault-tolerant gate sets~\cite{Beverland2020}.

The methods presented in this paper contribute to the existing repertoire of approximation techniques.
Having a variety of approximation methods is valuable because different approaches offer different space-time tradeoffs when approximating unitaries.
Depending on the specific constraints of a quantum computing architecture, such as the availability of ancillary qubits, the relative costs of different gate
types, or requirements on circuit depth, one approximation method may be preferable to another.

A key feature of our approach is that we directly approximate general one-qubit unitaries in $U(2)$, 
thereby avoiding the need for Euler angle decomposition or the magnitude approximation problem~\cite{KLMPP2023}.
Traditional approaches typically decompose an arbitrary one-qubit unitary into a product of rotations about fixed axes,
and then approximate each rotation separately. Our method circumvents this intermediate step entirely.

Recent work has explored practical direct $U(2)$ approximations using different techniques.
Morisaki et al.~\cite{morisaki2025} and Hao et al.~\cite{hao2025} have developed ancilla-free methods for direct unitary approximation,
including approaches based on mixing. Ancilla-free methods have the advantage of not requiring additional qubits beyond those used by the target unitary.
In contrast, our approach requires one ancillary qubit but might offer complementary benefits such as shorter circuit depth or lower gate counts,
with high probability.

Approximations with orthogonal gate sets and multi-qubit gate sets have been enabled through catalysis techniques~\cite{amy2024}.
Our work provides an alternative approach to achieving similar goals.
Rather than relying on catalysis, we use repeat-until-success circuits combined
with lattice-based exact synthesis algorithms to handle a variety of gate sets, including those based on orthogonal matrices
and multi-qubit gates such as Clifford and CS, Clifford and CCZ, and Real Clifford and CCZ.

A more extensive discussion of prior work on approximate synthesis published before 2023
can be found in the introduction to~\cite{KLMPP2023}.

The remainder of this paper is organized as follows. 
In~\cref{sec:key-example}, we present the key example of our approximation technique using Clifford and CS gates.
We then discuss generalizations to different notions of approximation,
approximations over real numbers, and higher degree number fields.
Finally, we conclude with an outlook on future work.

\section{Preliminaries and notation}

Here we review some well-known results and establish notation.

\subsection{Common unitary matrices}

Recall that one-qubit Pauli matrices are
\begin{equation}
I = 
\left(
\begin{array}{cc}
    1 & 0 \\
    0 & 1
\end{array}
\right),\,
X = 
\left(
\begin{array}{cc}
    0 & 1 \\
    1 & 0
\end{array}
\right),\,
Y =
\left(
\begin{array}{cc}
    0 & -i \\
    i & 0
\end{array}
\right),\,
Z =
\left(
\begin{array}{cc}
    1 & 0 \\
    0 & -1
\end{array}
\right),\,
\end{equation}
Clifford unitaries $C$ map tensor products of Pauli matrices to signed tensor products of Pauli matrices: 
$$
  C \{I,X,Y,Z\}^{\otimes n} C^\dagger \subset \pm\{ I,X,Y,Z\}^{\otimes  n}
$$
Other common matrices used in quantum computing are:
\begin{equation}
\label{eq:common-complex}
\text{S}=\left(\begin{array}{cc}
1 & 0\\
0 & i
\end{array}\right),\,\text{T}=\left(\begin{array}{cc}
1 & 0\\
0 & \zeta_{8}
\end{array}\right),\,\sqrt{\text{T}}=\left(\begin{array}{cc}
1 & 0\\
0 & \zeta_{16}
\end{array}\right)
\end{equation}

\begin{equation}
\label{eq:common-real}
\text{S}_{y} = \left(\begin{array}{cc}
\cos(\frac{\pi}{4}) & -\sin(\frac{\pi}{4})\\
\sin(\frac{\pi}{4}) & \cos(\frac{\pi}{4})
\end{array}\right),\,\text{T}_{y}=\left(\begin{array}{cc}
\cos(\frac{\pi}{8}) & -\sin(\frac{\pi}{8})\\
\sin(\frac{\pi}{8}) & \cos(\frac{\pi}{8})
\end{array}\right),\,\text{H}=\frac{1}{\sqrt{2}}\left(\begin{array}{cc}
1 & 1\\
1 & -1
\end{array}\right)
\end{equation}
For any $n$-qubit unitary $U$ we define $(n+1)$-qubit controlled-$U$
as 
\[
\text{C}U=\frac{I+Z}{2}\otimes I^{\otimes n} +\frac{I-Z}{2}\otimes U
\]
For example, we use matrices $\text{CT}$,$\text{CS}$, $\text{CH}$,
$\text{CT}_{y}$,$\text{CS}_{y}$, $\text{CCS}_{y}$, $\text{CCS}$, $\text{CCZ}$ when defining gate sets. 

\subsection{Unitaries and linear combinations of Pauli matrices}

Recall that any two-by-two complex matrix $U$ with real determinant can be associated with a four-dimensional 
real vector $u = (u_I,u_X,u_Y,u_Z)$ as follows 
\begin{equation} \label{eq:vector-matrix}
    U = u_I \cdot I + u_X \cdot i X + u_Y \cdot i Y + u_Z \cdot i Z
\end{equation}
This correspondence has the following useful properties: 
\begin{align}
    UU^\dagger = \ip{u,u}\cdot I,\quad \Tr(UV^\dagger) = 2\ip{u,v}, \quad \det U = \ip{u,u},  
\end{align}
where $\ip{u,v}$ is the standard Euclidean inner product and $\nrm{u}=\sqrt{\ip{u,u}}$.
With the standard matrix inner product and $2$-norm defined as 
\begin{equation}\label{eq:matrix-ip-norm}
    \ip{U,V} = \Tr(UV^\dagger), \quad \nrm{U}_2 = \sqrt{\ip{U,U}}
\end{equation}
the norms and inner products of matrices and corresponding real vectors are related as: 
\begin{equation} \label{eq:matrix-vector-ip}
    \ip{U,V} = 2\ip{u,v},\quad \nrm{U}_2 = \sqrt{2} \nrm{u}.
\end{equation}
When $u$ is a unit vector, the corresponding matrix is an element of the special unitary group, that is, a unitary with determinant one. Similarly, the rescaled matrix $\frac{U}{\nrm{u}}$ is always unitary for any non-zero vector $u$.
\subsection{Isometries as sub-matrices of unitaries} 

Recall that a complex matrix is an isometry if its columns have norm one and are mutually orthogonal. 
In particular, restricting a unitary matrix to a subset of its columns yields an isometry.

\subsection{Repeat-until-success circuits} 

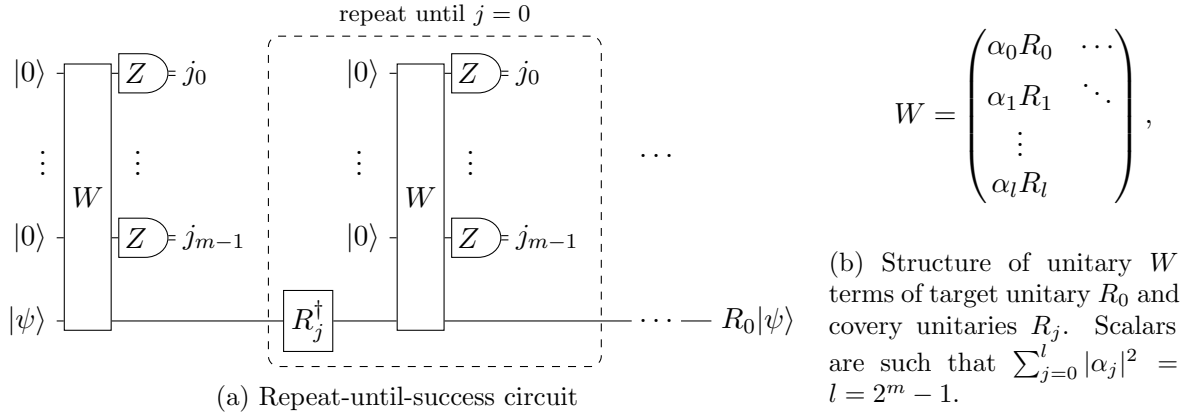
\begin{figure}[htbp]
\centering
\begin{subfigure}[c]{0.65\textwidth}
\centering
\begin{tikzpicture}
\begin{yquant}[register/separation=4mm]
    qubit {$\ket{0}$} a1;
    qubit {$\vdots$} d1;
    discard d1;
    qubit {$\ket{0}$} a2;
    
    qubit {$\ket{\psi}$} psi;
    
    box {$W$} (a1, d1, a2, psi);
    
    dmeter {$Z$} a1;
    text {$j_0$} a1;
    text {$\vdots$} d1;
    dmeter {$Z$} a2;
    text {$j_{m-1}$} a2;
    discard a1;
    discard a2;

    align a1, d1, a2, psi;

    hspace {3mm} psi;
    [name=recover]
    box {$R_j^\dagger$} psi;

    align psi, a1, d1, a2;

    [name=i1]
    init {$\ket{0}$} a1;
    text {$\vdots$} d1;
    [name=i2]
    init {$\ket{0}$} a2;

    [name=w2]
    box {$W$} (a1, d1, a2, psi);
    
    [name=meas1]
    dmeter {$Z$} a1;
    [name=meas1]
    text {$j_0$} a1;
    text {$\vdots$} d1;
    dmeter {$Z$} a2;
    [name=meas2]
    text {$j_{m-1}$} a2;
    
    discard a1;
    discard d1;
    discard a2;

    align psi, a1, d1, a2;
    hspace {5mm} psi;
    align psi, a1, d1, a2;
    text {$\hdots$} d1;
    text {$\hdots$} psi;
    hspace {3mm} psi;
    output {$R_0 \ket{\psi}$} psi;

    \node[draw, dashed,rounded corners, inner xsep = 2mm, inner ysep = 2mm, fit=(recover) (i1) (i2) (meas1) (meas2) ,label={[align=center]above:{\footnotesize repeat until $j = 0$}}] {};

\end{yquant}

\end{tikzpicture}
\caption{Repeat-until-success circuit}
\label{fig:rus-circuit-diagram}
\end{subfigure}%
\hfill
\begin{subfigure}[c]{0.32\textwidth}
\centering
\begin{align*}
    W = & 
\begin{pmatrix}
\alpha_0 R_0 & \cdots \\
\alpha_1 R_1 & \ddots \\
\vdots & \\
\alpha_l R_l &
\end{pmatrix},
\end{align*}
\caption{Structure of unitary $W$ in terms of target unitary $R_0$ and recovery unitaries $R_j$.
Scalars $\alpha_j$ are such that $\sum_{j = 0}^{l} |\alpha_j|^2 = 1$, $l = 2^{m}-1$.
}
\label{fig:rus-isometry}
\end{subfigure}
\caption{Repeat-until-success circuit that implements one-qubit unitary $R_0$~(Figure~2 in \cite{RUS-Paetznick-Svore}). Outcomes $j_0, \ldots, j_{m-1}$ are the bits of the $m$-bit integer $j$.
When $j$ is zero, the circuit succeeds and implements one-qubit unitary $R_0$.
The success probability is $|\alpha_0|^2$.
When $j$ is non-zero, the circuit fails and 
the recovery unitary $R_j$ ($j > 0$) must be applied, followed by another iteration of the circuit.}
\label{fig:rus-circuit}
\end{figure}

The key ingredient in our approximation algorithm is the repeat-until-success circuit construction introduced in~\cite{RUS-Paetznick-Svore}. 
We summarize this construction in~\cref{fig:rus-circuit}. 
Reference~\cite{RUS-Paetznick-Svore} explores circuits where the unitary $W$ comes from various families of multi-qubit Clifford and T circuits. In this work, we instead directly construct the first two columns of $W$ 
and then use recent exact synthesis algorithms~\cite{KS2024} for isometries to find a circuit for $W$ over various gate sets.

\subsection{Exact synthesis of isometries}
\label{sec:exact-synthesis}

Exact synthesis for isometries is a generalization of the exact synthesis problem for unitaries. 
The isometry synthesis problem of interest here is the following:
Given a $2^n \times 2$ isometry $W'$ with entries in a ring $\mathcal{R}_\xi$, find 
a circuit over the gate set corresponding to the ring~(see~\cref{tab:common-gate-sets}), such that the unitary $W$ implemented by 
the circuit has its first two columns equal to $W'$.
For exact synthesis of isometries, we rely on recent algorithms from~\cite{KS2024}, an instantiation 
of $A^\ast$ search with an appropriately chosen heuristic function. 
The exact synthesis problem for isometries described above is a special case of Problem~2.1 in~\cite{KS2024}.

\begin{table}[ht]
    \centering
\begin{tabular}{|c|c|c|c|c|}
\hline 
Gate set & Field $E$ & Denominator $\xi$ & Ring $\mathcal{R}_\xi$ & Minimal $n$\tabularnewline
\hline 
\hline 
Clifford and $\text{T}$  & $\mathbb{Q}(\zeta_{8})$ & $1+\zeta_{8}$ & $\mathbb{Z}[\zeta_8, 1/\xi]$ & $1$\tabularnewline
\hline 
Clifford and $\text{T}$, $\sqrt{\text{T}}$ & $\mathbb{Q}(\zeta_{16})$ & $1+\zeta_{16}$ & $\mathbb{Z}[\zeta_{16}, 1/\xi]$ &  $1$\tabularnewline
\hline 
\hline 
Clifford and $\text{CS}$ & $\mathbb{Q}(i)$ & $1+i$ &$\mathbb{Z}[i, 1/\xi]$ &  $2$\tabularnewline
\hline 
Clifford and $\text{CH}$ & $\mathbb{Q}(\sqrt{2})$ & $\sqrt{2}$ &$\mathbb{Z}[\sqrt{2}, 1/\xi]$ &  $2$\tabularnewline
\hline 
Clifford and $\text{T},\text{T}\otimes\text{T},\text{CT}$ & $\mathbb{Q}(\zeta_{8})$ & $1+\zeta_{8}$ & $\mathbb{Z}[\zeta_8, 1/\xi]$ & $2$\tabularnewline
\hline 
Clifford and $\text{T}_{y},\text{CS}_{y},\text{CT}_{y}$ & $\mathbb{Q}(\cos(\frac{\pi}{8}))$ & $2+2\cos(\frac{\pi}{8})$ &$\mathbb{Z}[\xi, 1/\xi]$ &  $2$\tabularnewline
\hline 
Clifford and $\text{T},\text{T}^{\otimes2},\ldots,\text{T}^{\otimes n}$ & $\mathbb{Q}(\zeta_{8})$ & $1+\zeta_{8}$ &$\mathbb{Z}[\zeta_8, 1/\xi]$ &  $2$\tabularnewline
\hline 
\hline 
Clifford and $\text{CS}$, $\text{CCZ}$, $\text{CCS}$ & $\mathbb{Q}(i)$ & $1+i$ &$\mathbb{Z}[i, 1/\xi]$ &  $3$\tabularnewline
\hline 
Clifford and $\text{CS}_{y}$, $\text{CCZ}$, $\text{CCS}_{y}$ & $\mathbb{Q}(\sqrt{2})$ & $\sqrt{2}$ & $\mathbb{Z}[\sqrt2, 1/\xi]$ & $3$\tabularnewline
\hline 
\end{tabular}
    \caption{Common $n$-qubit gate sets and associated rings $\mathcal{R}_\xi$. The approximation algorithms considered in this paper are compatible with all of these gate sets. We omit the word ``real'' in front of ``Clifford'' in the gate set name, as it is implied by the corresponding field $E$ being totally real.
    We use $\zeta_m = e^{i \frac{2\pi}{m}}$ for the principal $m$-th root of unity.
    \label{tab:common-gate-sets}
    }
\end{table}

Before applying the search, we need to establish that a solution to the exact synthesis problem exists. Fortunately, there is a vast literature on 
this topic~\cite{Giles2013,Amy2020,amy2024} that applies to exact synthesis of isometries as well as unitaries.

\section{Approximating unitaries via repeat-until-success circuits}

We show how to use repeat-until-success circuits, as defined in~\cite{RUS-Paetznick-Svore}, to approximate
arbitrary two-by-two unitaries, bypassing the need for Euler angle decomposition or similar techniques such as magnitude approximation~\cite{KLMPP2023}.  

\subsection{Key example: approximation with Clifford and CS gates} \label{sec:key-example}

The input to our approximation problem is a target one-qubit special unitary $U$, an accuracy $\varepsilon$, and a maximum failure probability $p$
beyond which the repeat-until-success circuit requires another iteration. 

Our goal is to construct a repeat-until-success circuit (\cref{fig:rus-circuit}), which we call a $(p,\varepsilon)$-approximation circuit for $U$, that has the following properties: 
\begin{itemize}
    \item the circuit uses one ancillary qubit ($m=1$ in \cref{fig:rus-circuit}),
    \item the success probability of the first step is at least $1-p$,
    \item upon success, the circuit implements a unitary $R_0$ such that $\nrm{U-R_0}_2 \le \varepsilon$,
    \item the unitary $W$ can be expressed as a circuit over the Clifford and CS gate set.
\end{itemize}

Upon failure of the repeat-until-success circuit, we have several options. We can either approximate the new target unitary $U R_1^\dagger$ using the same strategy, or fall back to a different approximation method that uses either Euler decomposition or magnitude approximation.
For this reason, we focus on constructing unitary $W$ used in the first step of the repeat-until-success circuit.

\begin{enumerate}
    \item \textbf{[Point enumeration]} Set $N=2$, find vector $u_0$ with integer coordinates such that: 
    \begin{itemize}
        \item Norm squared of $u_0$ satisfies: $(1-p) 2^N \le \ip{u_0, u_0} \le 2^N$,
        \item The corresponding matrix $U_0$~(\cref{eq:vector-matrix}) satisfies:  $\nrm{ \frac{U_0}{\nrm{u_0}} - U }_2 \le \varepsilon $.
    \end{itemize}
    If there are no such vectors $u_0$, increase $N$ by 1 and repeat this step.
    \item \textbf{[Norm equation]} Find an integer vector $u_1$ such that $ \ip{u_1, u_1} = 2^N - \ip{u_0, u_0}$ 
    \item \textbf{[Exact synthesis]} Use exact synthesis for isometries to find a circuit that implements a $4\times 4$ unitary $W = W_{N,u_0,u_1}$~(\cref{fig:rus-circuit}) below
    over the Clifford and CS gate set.
    $$
     W_{N,u_0,u_1} = 
        \begin{pmatrix}
        \alpha_0 R_0 & \ast \\
        \alpha_1 R_1 & \ast \\
        \end{pmatrix} 
       = 
        \begin{pmatrix}
        \frac{U_0}{(1+i)^N} & \ast \\
        \frac{U_1}{(1+i)^N} & \ast \\
        \end{pmatrix} 
    $$
    given its first two columns. Unitaries $U_0, U_1$ correspond to vectors $u_0, u_1$ via \cref{eq:vector-matrix}.
\end{enumerate}

The algorithm above finds a $(p,\varepsilon)$-approximation circuit for $U$ because, in the notation of \cref{fig:rus-circuit}, $R_0 = U_0 / \nrm{u_0}$. This ensures that $\nrm{U-R_0}_2 \le \varepsilon$.
The equality $\alpha_0 = \nrm{u_0}/(1+i)^N$ implies that the success probability is 
$\ip{u_0,u_0} / 2^N$, which ensures the success probability is at least $1-p$.

The three steps above are the standard steps in any number-theoretic unitary approximation technique. 
Next, we expand on the details of each step.

\subsubsection{Point enumeration}

We show that the following point enumeration problem can be reduced to enumerating integer points 
in a convex body, similar to how a non-convex constraint is strengthened when approximating diagonal one-qubit unitaries by circuits with fall-back~\cite{KLMPP2023}.

\begin{problem}[Point enumeration]\label{prob:point-enumeration}
Given $N$, $p \in (0,1)$, a special unitary $U$, and $2 > \varepsilon > 0$, find $u_0 \in \mathbb{Z}^4$ such that 
$(1-p) 2^N \le \ip{u_0, u_0} \le 2^N$ and $\nrm{ \frac{U_0}{\nrm{u_0}} - U }_2 \le \varepsilon$,
where the matrix $U_0$ corresponds to $u_0$ via \cref{eq:vector-matrix}.
\end{problem}

Define $u$ to be a unit vector corresponding to unitary $U$ via~\cref{eq:vector-matrix}. 
We show that $\nrm{ \frac{U_0}{\nrm{u_0}} - U }_2 \le \varepsilon $ is a convex constraint and 
strengthen $(1-p) 2^N \le \ip{u_0, u_0}$ into a convex one.

First, using \cref{eq:matrix-ip-norm,eq:matrix-vector-ip}, inequality $\nrm{ \frac{U_0}{\nrm{u_0}} - U }_2 \le \varepsilon $ is equivalent to
$$
\left\langle \frac{u_0}{\nrm{u_0}}, u \right\rangle \ge 1 - \left(\frac{\varepsilon}{2}\right)^2
$$
The above constraint on $u_0$ is convex. Using $c = 1 - \left(\frac{\varepsilon}{2}\right)^2$,
we rewrite it as 
\begin{equation}\label{eq:epsilon-constraint}
 \ip{ u, u_0}^2 \ge c^2 \ip{ u_0, u_0},
\end{equation}
which is an example of a hyperbolic cone (see Example~2.11 in~\cite{BoydVandenberghe2004}).

Second, we use $\ip{u_0, u_0} \ge |\ip{u_0, u}|^2 $ and strengthen  $(1-p) 2^N \le \ip{u_0, u_0}$
to 
\begin{equation}\label{eq:probability-constraint}
|\ip{u_0, u}| \ge \sqrt{(1-p)\cdot 2^N}
\end{equation}
The constraints in \cref{eq:epsilon-constraint,eq:probability-constraint}, together with $\ip{u_0,u_0} \le 2^N$,
define a convex region $\mathcal{R}_{u,\varepsilon,p,N}$.
We have reduced the point enumeration problem (\cref{prob:point-enumeration}) to enumerating integer points in a convex region,
which can be solved in polynomial time.

\subsubsection{Norm equation}

The norm equation step requires finding four integers $u_{1,I}, u_{1,X}, u_{1,Y}, u_{1,Z}$ such that 
$$
u_{1,I}^2 +  u_{1,X}^2  + u_{1,Y}^2 + u_{1,Z}^2 = 2^N - \ip{u_0,u_0}.
$$
A solution to such an equation always exists. This fact is known as the four squares theorem in number theory.
Moreover, there are efficient randomized algorithms for this problem; see the recent review~\cite{Pollack2018}.
The reason we call this a norm equation is that this problem can be thought of as finding a quaternion in the Lipschitz order in the quaternion algebra $\left( \frac{-1,-1}{\mathbb{Q}}\right)$. 
Relating this problem to quaternions has a couple of advantages: it gives insight into all possible solutions 
to the four squares equation and is crucial for generalizing our approximation strategy to other gate sets.

It can be beneficial to consider several solutions, as some of them might lead to a shorter 
circuit needed to implement unitary $W$ in the repeat-until-success circuit.

\subsubsection{Exact synthesis}

To find the circuit for $W$, we use the exact synthesis algorithm discussed in~\cref{sec:exact-synthesis}.
This is possible because we have constructed the first two columns of $W$ as a $4\times2$ isometry with entries in 
$\mathbb{Z}[i,1/(1+i)]$. 
The integer $N$ controls the length of the circuit; however, we do not have a tight correspondence between the number of 
$\text{CS}$ gates in the circuit and $N$, as is the case for the one-qubit Clifford and T gate set.
For this reason, it can be beneficial to consider many solutions in the point enumeration and norm equation steps of the algorithm for varying $N$ and then pick the best circuit for $W$. 
Our application of exact synthesis algorithms highlights a need 
for exact synthesis for isometries with stronger theoretical guarantees on the relation between $N$
and the circuit's gate count.

\subsection{Generalizations} 

There are three main directions in which our key example can be generalized: 
\begin{itemize}
    \item Different notions of approximation. For example, ensuring that the target $U$ is approximated within diamond norm distance $\varepsilon$, or that the target $U$ is approximated as a mixture of unitary channels.
    \item Approximations over real numbers. It may be interesting to consider approximating orthogonal $U$ using gate sets consisting of orthogonal matrices, that is, using real numbers as opposed to complex numbers.
    \item Higher degree number fields. Many interesting gate sets are related to higher-degree number fields, for example $\mathbb{Q}(\zeta_8)$ for Clifford and $T$, 
    as opposed to $\mathbb{Q}(i)$.
\end{itemize}

We now discuss each of these generalizations in more detail.

\subsubsection{Different notions of approximation}

Different notions of approximation lead to different point enumeration problems.
This is analogous to how different point enumeration problems arise in approximating 
one-qubit diagonal unitaries in~\cite{KLMPP2023}. 
A strategy for mixed approximation of arbitrary $U(2)$ is outlined in~\cite{Campbell2017}.
Similar to~\cite{KLMPP2023}, this strategy can be more deeply integrated with the approximation algorithm 
described here by modifying the point enumeration step. 
If the target unitary $U$ is diagonal, Clifford twirling can be used to simplify the calculation 
of the diamond distance between target unitary $U$ and the approximating mixture.

\subsubsection{Approximations over real numbers} \label{sec:real}

In some settings, one might be interested in working with a gate set consisting of orthogonal matrices 
as opposed to unitary ones. In this case, we can restrict target unitaries to orthogonal matrices 
$$
 U_\mathbb{R} = u_I I + u_Y iY
$$
by setting the coefficients $u_X, u_Z$ to zero. 
If we do not want to add an extra qubit, the norm equation step requires solving a two-squares equation,
which does not always have a solution. One can either iterate over various solutions to the point enumeration step until a solution exists, or add another qubit and find three recovery matrices $R_1, R_2, R_3$ by solving a six-squares equation, which always has a solution, similar to the four-squares case.

\subsubsection{Higher degree number fields}

Suppose that the gate set of interest is defined over a number field $E$ from~\cref{tab:common-gate-sets}, and 
let $F$ be its maximal totally real subfield. For example, when considering the Clifford and T gate set, 
$E = \mathbb{Q}(\zeta_8)$ and $F = \mathbb{Q}(\sqrt 2)$. We first discuss a simple generalization of the algorithm 
in~\cref{sec:key-example}. At the end of this section, we outline a more sophisticated generalization.

The simple generalization is to consider vectors $u_0, u_1$ with entries in $O_F$, the ring of integers of $F$.
This requires modifying the point enumeration and norm equation steps, and introducing additional notation to handle the general situation.
Recall that a totally real field $F$ of degree $d$ has $d$ different embeddings into $\mathbb{R}$, that is, 
maps that respect field operations. For example, $F=\mathbb{Q}(\sqrt{2})$ has two embeddings $\sigma_1, \sigma_2$: 
$$
 \sigma_1(q_1 + \theta q_2) = q_1 + \sqrt{2} q_2,\quad \sigma_2(q_1 + \theta q_2) = q_1 - \sqrt{2} q_2, \text{ for } q_1, q_2 \in \mathbb{Q},\,\theta \in \mathbb{Q}(\sqrt{2}),\,\theta^2 = 2,
$$
and the ring of integers $O_F$ is 
$$
\mathbb{Z}[\sqrt{2}] = \{ a_1 + \theta a_2 : a_1,a_2 \in \mathbb{Z} \}.
$$
We use $\sigma_k(u_0)$ to denote the real vector obtained by applying $\sigma_k$ to each coordinate of $u_0$,
and $\delta=|\xi|^2 \in F$ to denote the absolute value squared of the ring denominator $\xi$ in \cref{tab:common-gate-sets}.

The point enumeration step is modified as follows.
Because each element of $O_F$ can be represented using $d$ integers
$$
 \sum_{j=1}^d a_j z_j \text{ for } a_j \in \mathbb{Z}, \text{ where } z_j \in O_F \text{ is an integral basis of } O_F,
$$
the new point enumeration problem is concerned with finding integer points in a $4d$-dimensional convex body given by the following constraints: 
\begin{align}
\left\langle \frac{\sigma_1(u_0)}{\nrm{\sigma_1(u_0)}}, u \right\rangle 
& \ge
1 - \left(\frac{\varepsilon}{2}\right)^2 \\
|\ip{\sigma_1(u_0), u}| 
& \ge
\sqrt{(1-p)\cdot \sigma_1(\delta)^N} \\
\ip{\sigma_k(u_0),\sigma_k(u_0)} 
& \le
\sigma_k(\delta)^N, \text{ for } k = 1,\ldots,d
\end{align}
The last set of constraints ensures that the norm equation 
$$
 \ip{u_1,u_1} = \delta^N - \ip{u_0,u_0}
$$
has a positive right-hand side for all embeddings into $\mathbb{R}$, which is necessary 
for the existence of a solution. 
This is directly analogous to how point enumeration problems are set up for various approximation questions considered in~\cite{KLMPP2023}.

To solve the generalized norm equation 
$$
u_{1,I}^2 +  u_{1,X}^2  + u_{1,Y}^2 + u_{1,Z}^2 = \delta^N - \ip{u_0,u_0},
$$
one can rely on the techniques for solving relative norm equations used for 
approximate circuit synthesis in~\cite{KLMPP2023}. 
We first randomly pick $u_{1,I}$ and $u_{1,X}$ such that 
$$
\sigma_k(u_{1,I}^2 +  u_{1,X}^2 ) \le \sigma_k(\delta^N - \ip{u_0,u_0}), \text{ for } k = 1,\ldots,d
$$
and then solve equation 
$$
 u_{1,Y}^2 + u_{1,Z}^2 = \delta^N - \ip{u_0,u_0} - u_{1,I}^2 - u_{1,X}^2
$$
which can be solved as a relative norm equation.
This is a direct generalization of one of the randomized algorithms for finding a four-squares representation 
of a positive integer.

When the number field $E$ is totally real, we can follow the same strategies as in~\cref{sec:real}.
We set $u_{0,X}, u_{0,Z}, u_{1,X}, u_{1,Z}$ to zero and optionally add another qubit to ensure that 
a solution to the point enumeration problem can always be completed to an isometry. 

One can potentially get slightly better approximations by relying on maximal orders 
in quaternion algebras. The set 
$$
\{ u_I \cdot I + u_X \cdot i X + u_Y \cdot i Y + u_Z \cdot i Z : u_I,u_X,u_Y,u_Z \in O_F \}
$$
corresponds to an order in a quaternion algebra $\left(\frac{-1,-1}{F}\right)$.
Using a maximal order that includes it can lead to a finer grid for integer point enumeration 
and to a cleaner setup for the norm equation part. However, given that we do not have a tight relation between $N$, 
the power of $\delta$, and the cost of circuits needed to implement $W$, this question can only be explored numerically. 

\section{Outlook and future work}

We have outlined a theoretical basis for a family of new approximation algorithms. It will be interesting to implement these algorithms in a variety of settings and compare them to existing alternatives. Due to the nature of these algorithms, their performance and circuit cost scaling with the approximation parameters $p$ and $\varepsilon$ can only be determined through numerical study. 

\section*{Acknowledgment}
VK used GitHub Copilot with Claude 4.5 for drafting the introduction and proofreading this paper.
This work was completed while V.K. was a researcher at Microsoft Quantum. 
J.B. worked on this project during his internship at Microsoft Research during summer of 2018.
 
\bibliographystyle{plainurl}
\bibliography{references}

\end{document}